\documentclass[12pt]{article}
\usepackage{amsmath,amssymb,amsthm}

\theoremstyle{definition}

\theoremstyle{remark}

\newcommand{\beq}{\begin{eqnarray}}
\newcommand{\eeq}{\end{eqnarray}}
\newcommand{\beqnn}{\begin{eqnarray*}}
\newcommand{\eeqnn}{\end{eqnarray*}}
\newcommand{\rd}{\partial}

\newcommand{\ZZ}{\mathbf{Z}}
\newcommand{\gl}{\mathrm{gl}}
\newcommand{\GL}{\mathrm{GL}}
\newcommand{\bst}{\boldsymbol{t}}
\newcommand{\bsT}{\boldsymbol{T}}
\newcommand{\bszero}{\boldsymbol{0}}

\begin{document}

\title{Toda tau functions with quantum torus symmetries}
\author{Kanehisa Takasaki%
\thanks{E-mail: takasaki@math.h.kyoto-u.ac.jp}\\
{\small
Graduate School of Human and Environmental Studies,
Kyoto University}\\
{\small Yoshida, Sakyo, Kyoto, 606-8501, Japan}}
\date{}
\maketitle

\begin{abstract}
The quantum torus algebra plays an important role 
in a special class of solutions of the Toda hierarchy.  
Typical examples are the solutions related to 
the melting crystal model of topological strings 
and 5D SUSY gauge theories. 
The quantum torus algebra is realized 
by a 2D complex free fermion system that underlies 
the Toda hierarchy, and exhibits mysterious 
``shift symmetries''.  This article is based on 
collaboration with Toshio Nakatsu. 
\end{abstract}

\begin{flushleft}
\end{flushleft}

\section{Introduction}

This paper is a review of our recent work \cite{NT07,NT08} 
on an integrable structure of the melting crystal model of 
topological strings \cite{ORV03} and 5D gauge theories \cite{MNTT04}.  
It is shown here that the partition function of this model, 
on being suitably deformed by special external potentials, 
is essentially a tau function of the Toda hierarchy \cite{TT95}. 
A technical clue to this observation is a kind of symmetries 
(referred to as ``shift symmetries'') in the underlying 
quantum torus algebra.  These symmetries enable us, 
firstly, to convert the deformed partition function 
to the tau function and, secondly, to show the existence 
of hidden symmetries of the tau function.  These results 
can be exntended to some other Toda tau functions 
that are related to the topological vertex \cite{Zhou03}
 and the double Hurwitz numbers of the Riemann sphere 
\cite{Okounkov00}.

\section{Quantum torus algebra}

Throughout this paper, $q$ denotes a constant with $|q| < 1$, 
and $\Lambda$ and $\Delta$ denote the $\ZZ \times \ZZ$ matrices 
\beqnn
  \Lambda = \sum_{i\in\ZZ}E_{i,i+1} = (\delta_{i+1,j}), \quad 
  \Delta = \sum_{i\in\ZZ}iE_{ii} = (i\delta_{ij}). 
\eeqnn
Their combinations 
\beq
  v^{(k)}_m = q^{-km/2}\Lambda^m q^{k\Delta}
  \quad (k,m \in \ZZ)  
\eeq
satisfy the commutation relations 
\beq
  [v^{(k)}_m,\, v^{(l)}_n] 
  = (q^{(lm-kn)/2} - q^{(kn-lm)/2})v^{(k+l)}_{m+n}
\eeq
of the quantum torus algebra.  This Lie algebra 
can thus be embedded into the Lie algebra $\gl(\infty)$ 
of $\ZZ\times\ZZ$ matrices $A = (a_{ij})$ for which 
$\exists N$ such that $a_{ij} = 0$ for $|i-j| > N$. 

To formulate a fermionic realization of this Lie algebra, 
we introduce the creation/annihilation operators 
$\psi_i,\psi^*_i$ ($i \in \ZZ$) with anti-commutation relations 
\beqnn
  \psi_i\psi^*_j + \psi^*_j\psi_i = \delta_{i+j,0}, \quad
  \psi_i\psi_j + \psi_j\psi_i = 0, \quad 
  \psi^*_i\psi^*_j+\psi^*_j\psi^*_i = 0 
\eeqnn
and the 2D free fermion fields 
\beqnn
  \psi(z) = \sum_{i\in\ZZ}\psi_iz^{-i-1},\quad 
  \psi^*(z) = \sum_{i\in\ZZ}\psi^*_iz^{-i}.  
\eeqnn
The vacuum states $\langle 0|$, $|0\rangle$ 
of the Fock space and its dual space 
are characterized by the vacuum conditions 
\beqnn
\begin{aligned}
  &\psi_i|0\rangle = 0 \; (i \ge 0),&
  &\psi^*_i|0\rangle = 0\; (i \ge 1),\\
  &\langle 0|\psi_i = 0 \; (i \le -1),&
  &\langle 0|\psi^*_i = 0 \; (i \le 0). 
\end{aligned}
\eeqnn

To any element  $A = (a_{ij})$ of $\gl(\infty)$, 
one can associate the fermion bilinear 
\beqnn
  \widehat{A} = \sum_{i,j\in\ZZ}a_{ij}{:}\psi_{-i}\psi^*_j{:},\quad 
  {:}\psi_{-i}\psi^*_j{:} 
  = \psi_{-i}\psi^*_j - \langle 0|\psi_{-i}\psi^*_j|0\rangle. 
\eeqnn
These fermion bilinears form a one-dimensional central extension 
$\widehat{\gl(\infty)}$ of $\gl(\infty)$.  
The special fermion bilinears \cite{NT07,NT08} 
\beq
  V^{(k)}_m 
  = \widehat{v^{(k)}_m} 
  = q^{k/2}\oint\frac{dz}{2\pi i}
     z^m {:}\psi(q^{k/2}z)\psi^*(q^{-k/2}z){:} 
\eeq
satisfy the commutation relations 
\beq
 [V^{(k)}_m,\, V^{(l)}_n] 
 = (q^{(lm-kn)/2} - q^{(kn-lm)/2})
   \left(V^{(k+l)}_{m+n} - \frac{q^{k+l}}{1-q^{k+l}} \delta_{m+n,0}\right) 
\eeq
for $k$ and $l$ with $k + l \not= 0$ and 
\beq
   [V^{(k)}_m,\, V^{(-k)}_n] 
 = (q^{-k(m+n)/2} - q^{k(m+n)/2})V^{(0)}_{m+n} + m\delta_{m+n,0}. 
\eeq
Thus $\widehat{\gl(\infty)}$ contains a central extension 
of the quantum torus algebra, in which 
the $\widehat{u(1)}$ algebra is realized by 
\beq
  J_m = V^{(0)}_m = \widehat{\Lambda^m} \quad(m \in \ZZ). 
\eeq

\section{Shift symmetries}

Let us introduce the operators 
\beq
  G_{\pm} = \exp\left(\sum_{k=1}^\infty 
            \frac{q^{k/2}}{k(1-q^k)}J_{\pm k}\right), \quad
  W_0 = \sum_{n\in\ZZ}n^2{:}\psi_{-n}\psi^*_n{:}. 
\eeq
$G_{\pm}$'s play the role of ``transfer matrices'' 
in the melting crystal model\cite{ORV03,MNTT04}.  
$W_0$ is a fermionic form of the so called 
``cut-and-join'' operator for Hurwitz numbers 
\cite{Kazarian08}.  

$G_{\pm}$ and $q^{W_0/2}$ induce the following two types 
of ``shift symmetries'' \cite{NT07,NT08} 
in the (centrally extended) quantum torus algebra. 
\begin{itemize}
\item First shift symmetry 
\begin{multline}
  G_{-}G_{+}\left(V^{(k)}_m - \delta_{m,0}\frac{q^k}{1-q^k}
  \right)(G_{-}G_{+})^{-1} \\
  = (-1)^k\left(V^{(k)}_{m+k} - \delta_{m+k,0}\frac{q^k}{1-q^k}\right) 
\label{first-ss}
\end{multline}
\item Second shift symmetry
\beq
  q^{W_0/2}V^{(k)}_mq^{-W_0/2} = V^{(k-m)}_m
\label{second-ss}
\eeq
\end{itemize}

\section{Toda tau function in melting crystal model}

A general tau function of the 2D Toda hierarchy \cite{TT95} 
is given by 
\beq
  \tau(s,\bsT,\bar{\bsT}) 
  = \langle s|\exp\left(\sum_{k=1}^\infty T_kJ_k\right)
    g \exp\left(- \sum_{k=1}^\infty \bar{T}_kJ_{-k}\right) 
    |s \rangle, 
\eeq
where $\bsT = (T_1,T_2,\cdots)$ and $\bar{\bsT} 
= (\bar{T}_1,\bar{T}_2,\cdots)$ are time variables 
of the Toda hierarchy, $\langle s|$ and $|s\rangle$ 
are the ground states 
\beqnn
  \langle s| = \langle-\infty|\cdots \psi^*_{s-1}\psi^*_s,\quad  
  |s\rangle = \psi_{-s}\psi_{-s+1}\cdots |-\infty\rangle
\eeqnn
in the charge-$s$ sector of the Fock space, and $g$ is 
an element of $\GL(\infty) = \exp\bigl(\gl(\infty)\bigr)$. 

On the other hand, the partition function $Z(Q,s,\bst)$ 
of the deformed melting crystal model \cite{NT07,NT08} 
can be cast into the apparently similar 
(but essentially different) form 
\beq
  Z(s,\bst) = \langle s|G_{+}e^{H(\bst)}Q^{L_0}G_{-}|s\rangle, 
\eeq
where $Q$ and $\bst = (t_1,t_2,\cdots)$ are coupling constants 
of the model, and $H(\bst)$ and $L_0$ the following operators: 
\beq
  H(\bst) = \sum_{k=1}^\infty t_kH_k,\quad H_k = V^{(k)}_0,\quad 
  L_0 = \sum_{n\in\ZZ} n{:}\psi_{-n}\psi^*_n{:}. 
\eeq

The shift symmetries (\ref{first-ss}) and (\ref{second-ss}) 
imply the operator identity 
\beqnn
  G_{+}e^{H(\bst)}G_{+}^{-1} 
  = \exp\left(\sum_{k=1}^\infty\frac{t_kq^k}{1-q^k}\right)
    G_{-}^{-1}q^{-W_0/2}
    \exp\left(\sum_{k=1}^\infty (-1)^kt_kJ_k\right)q^{W_0/2}G_{-}. 
\eeqnn
Inserting this identity and using the fact that 
\beqnn
  \langle s|G_{-}^{-1}q^{-W_0/2} = q^{-s(s+1)(2s+1)/12}\langle s|,\quad
  q^{-W_0/2}G_{+}^{-1}|s\rangle = q^{-s(s+1)(2s+1)/12}|s\rangle,
\eeqnn
we can rewrite $Z(s,\bst)$ as 
\beq
  Z(Q,s,\bst) 
  = \exp\Bigl(\sum_{k=1}^\infty\frac{t_kq^k}{1-q^k}\Bigr) 
  q^{-s(s+1)(2s+1)/6} \tau(s,\bsT,\bszero), \quad 
    T_k = (-1)^kt_k, 
\eeq
where the $\GL(\infty)$ element $g$ defining 
the tau function is given by 
\beq
  g = q^{W_0/2}G_{-}G_{+}Q^{L_0}G_{-}G_{+}q^{W_0/2}. 
\eeq

Actually, the shift symmetries imply the operator identity 
\beqnn
  G_{-}^{-1}e^{H(\bst)}G_{-}
  = \exp\left(\sum_{k=1}^\infty\frac{t_kq^k}{1-q^k}\right)
    G_{+}q^{W_0/2}\exp\left(\sum_{k=1}^\infty (-1)^kt_kJ_{-k}\right)
    q^{-W_0/2}G_{+}^{-1} 
\eeqnn
as well.  This leads to another expression of $Z(Q,s,\bst,)$ 
in which $\tau(s,\bsT,\bszero)$ is replaced with 
$\tau(s,\bszero,-\bsT)$.  

The existence of different expressions can be explained 
by the intertwining relations 
\beq
  J_kg = gJ_{-k} \quad (k = 1,2,\ldots), 
\label{Jg=gJ}
\eeq
which, too, are a consequence of the shift symmetries.  
These intertwining relations imply the constraints 
\beq
  \left(\rd_{T_k} + \rd_{\bar{T}_k}\right)
  \tau(s,\bsT,\bar{\bsT}) = 0   \quad (k = 1,2,\ldots) 
\eeq
on the tau function.  The tau function $\tau(s,\bsT,\bar{\bsT})$ 
thereby becomes a function $\tau(s,\bsT-\bar{\bsT})$ 
of the difference $\bsT-\bar{\bsT}$.  In particular, 
$\tau(s,\bsT,\bszero)$ and $\tau(s,\bszero,-\bsT)$  coincide. 
The reduced function $\tau(\bsT,s)$ may be thought of 
as a tau function of the 1D Toda hierarchy.  

(\ref{Jg=gJ}) are a special case of the more general 
intertwining relations 
\beq
  (V^{(k)}_m - \delta_{m,0}\frac{q^k}{1-q^k})g 
  = Q^{-k}g(V^{(-k)}_{-2k-m} - \delta_{2k+m,0}\frac{q^{-k}}{1-q^{-k}}). 
\eeq
We can translate these relations to the language 
of the Lax formalism of the Toda hierarchy.  
A study on this issue is now in progress.

\section{Other models}

The following Toda tau functions can be treated more or less 
in the same way as the foregoing tau function.  
We shall discuss this issue elsewhere.  

\begin{itemize}
\item[1.] The generating function of the two-leg amplitude  
$W_{\lambda\mu}$ in the topological vertex \cite{Zhou03} 
is a Toda tau function determined by 
\beq
  g = q^{W_0/2}G_{+}G_{-}q^{W_0/2}. 
\eeq
\item[2.] The generating function of double Hurwitz numbers  
of the Riemann sphere \cite{Okounkov00} is a Toda tau function 
determined by 
\beq
  g = e^{-\beta W_0}Q^{L_0}. 
\eeq
The parameter $q$ is interpreted as $q = e^{-\beta}$.  
\end{itemize}

\subsection*{Acknowledgements}

This work has been partly supported by the JSPS Grants-in-Aid 
for Scientific Research No. 19104002, No. 21540218 
and No. 22540186 from the Japan Society 
for the Promotion of Science.

\end{document}